# Comsol Simulations of Cracking in Point Loaded Masonry with Randomly Distributed Material Properties

A.T. Vermeltfoort*[1], A.W.M. Van Schijndel[1]
[1]Eindhoven University of Technology, Dept. of the Built Environment
*P.O. Box. 513, 5600 MB Eindhoven, The Netherlands, a.t.vermeltfoort@tue.nl

**Abstract:** This paper describes COMSOL simulations of the stress and crack development in the area where a masonry wall supports a floor. In these simulations one of the main material properties of calcium silicate, its E-value, was assigned randomly to the finite elements of the modeled specimen. Calcium silicate is a frequently used building material with a relatively brittle fracture characteristic. Its initial E-value varies, as well as tensile strength and post peak behavior. Therefore, in the simulation, initial E-values were randomly assigned to the elements of the model and a step function used for describing the descending branch. The method also allows for variation in strength to be taken into account in future research. The performed non-linear simulation results are compared with experimental findings. They show the stress distribution and cracking behavior in point loaded masonry when varying material properties are used.

**Keywords:** Cracking, stress concentration, sequential linear analysis, random properties, step function

## 1. Introduction

Well designed masonry walls are capable to carry loads from floors and walls above. However, local contact effects may reduce the bearing capacity. One of these local contact situations occurs when a strip, smaller than the thickness of the wall, a so called centering strip, is placed on top of the wall. These centering strips are used when floors are relatively slender, i.e. floors with a large span-thickness ratio. Then it may be necessary to increase the rotation capacity of the floor-wall connection. Consequently, the load from the floor is concentrated. Figure 1 schematically shows floor-wall connections with and without a centering strip and the relating mechanical schemes in which the corner is either ridged (without strip) or hinged (with centering strip).

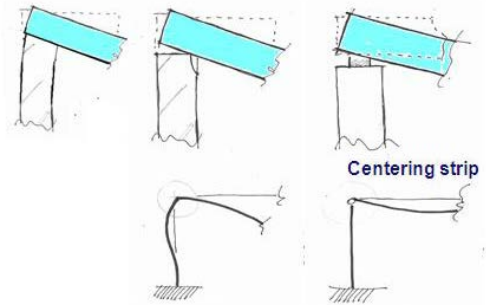

**Figure 1** Wall-floor connections and the use of a centering strip.

For a large rotation capacity (hinged connection) the width of the strip must be small, however, a small width causes higher splitting stresses in the wall.

Width, thickness and type of material of the strip play an important role in the deformation of the floor-wall connection. A flexible material with relatively much deformation at small loads is preferred because the effect of eccentric loading will be smaller compared to the use of a stiff material. Therefore, often rubber is used for centering strips. The ideal position of the strip is in the centre of the wall. In practice, the strips are sometimes positioned out of centre. This may affect the load bearing capacity.

To date, no explicit rules or design guidelines are available for the design and control of centering strip connections. Therefore, the work described in this paper was undertaken and finally, the effects of centre strips on the load bearing capacity of walls must become clear.

Firstly, this paper discusses the stress concentrations that occur in structures which are loaded on a relatively small area. This is followed, secondly, by a discussion of the experiments and its set up. The material, its construction features and its mechanical properties, like tensile strength (f'm) and modulus of elasticity (E) important for numerical simulations, are presented. A relationship between strip position and load bearing capacity is proposed. Thirdly, two kinds of simulations of cracking, using Comsol, are discussed and results compared with experiments.

It is concluded that the simulation of the non-linear behavior, using COMSOL, is satisfactory. Suggestions for the study of other parameter effects are given.

## 2 The CASIEL building system

Since the mid 1980s the use of CASIELs in medium rize buildings has become popular. The Calcium-Silicate industry has developed a complete program of building load-bearing walls, from design to finished walls, Berkers 1995 [2], Vermeltfoort and Ng'andu 2007 [11].

Compressive strength is one of the main properties used in the design of load bearing masonry. Other properties are related to it, like tensile strength.

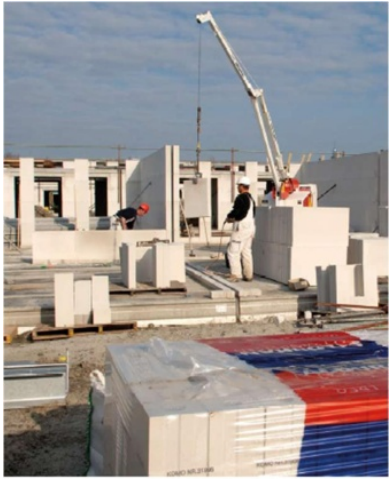

**Figure 2** Building with calcium silicate elements using a dedicated crane

Building with CASIELs has tremendously enhanced the speed of erecting walls while reducing labor costs and the physical stress of the bricklayers on site. Finishing costs are also significantly reduced due to the smoothness of the surface of CASIELs. Other factors cited in favor of calcium silicate elements include excellent structural performance of the material, environmental friendliness (the material can be crushed and reused) and better quality products due to the production of elements in factory controlled conditions. See also Figure 2.

## 3 Theory

### 3.1 Splitting strength

In areas where relatively large forces have to be transmitted via a relatively small contact surface stresses will become high. Studies, from e.g. Page [8], show that the material just below the connection is in a 3D compressive stress state. Further away, stresses fade out in the supporting wall. Consequently, at some distance from the support, the vertical compressive load causes tension in both horizontal directions, which, certainly in thickness direction, may lead to splitting of the wall. Of course, the magnitude of the stresses depends on the size of the contact area and its position on the wall, and the thickness of the wall. This kind of stress conditions also occur below centre strips.

The tensile strength of a material is one of the important structural design parameters. To establish the tensile strength a specimen can be loaded in tension. However, in brittle materials, like concrete, clay brick and calcium silicate (CaSi) it is not easy to apply the load to the specimen. Therefore, other test methods are developed, like the Brazilian splitting test, Blaauwendraad [3] and Heron [5].

In this splitting test, the load is introduced via two wooden laths of 5x5mm$^2$. Directly under these laths, the material is in a three dimensional stress situation. Further downwards, the lateral stresses in the vertical mid section are uniformly distributed. These tensile stresses are equal to:

$$\sigma = 2N/\pi.d.l \qquad (1)$$

with:

N = Load [kN]
d = specimen's height [mm]
l = specimen's length [mm]

For a cube sized specimen with rib length a, l and h are equal to a, ( d = l = a). Consequently, the strength (Fbu) equals:

$$fbu = 0.64 N/a^2 \qquad (2)$$

Splitting tests as described above were performed on the material used for the centering strip experiments that are discussed later in this paper.

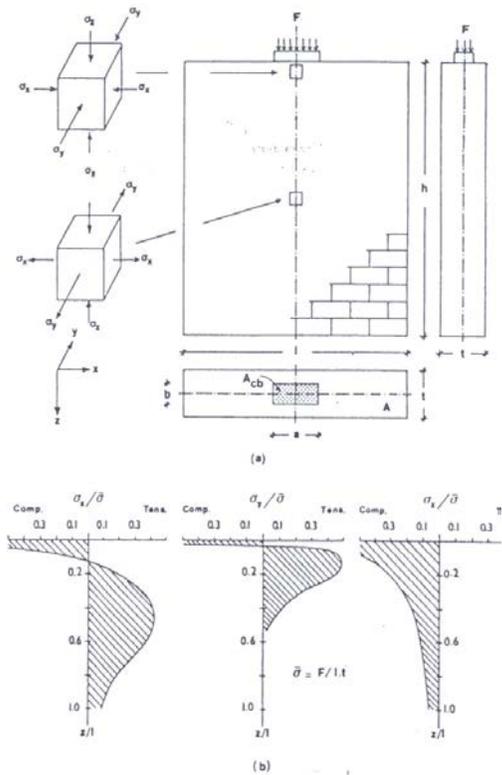

**Figure 3** Load-scheme for concentrated load on a wall, stress distribution in X, Y and Z direction respectively at vertical centre line of the load.

### 2.2 Specimen's dimension

As discussed earlier, only a relatively small part of a wall was studied. Therefore, blocks were cut form elements with a size big enough to allow for an area around the problem zone to spread the stresses resulting in a more or less uniform stress distribution at the opposite edge. Based on the Saint Venant principle, it was assumed that a height of the specimen of approximately twice the thickness of the wall would suffice. This is confirmed later in this paper.

The cut bottom surface was positioned on the bottom loading plate. The top edge simulated the wall-floor connection.

### 2.3 Stress-strain relationships and elasticity modulus of CASIEL

As mentioned earlier, compressive strength is one of the main properties used in the design of load bearing masonry but to establish the wind load distribution over the masonry structure, the modulus of elasticity is needed. For quality control purposes, the compressive strength (f'd) is determined from crushing tests on cubes or prisms cut out of calcium silicate (CaSi) elements at predetermined positions, EN 771-2:2003. The prisms are either dried before testing or a correction for moisture content is applied afterwards. Compressive strength of CaSi elements ranges from 16.0 to 36.0 Mpa with Young's moduli between 8000Mpa and 14000Mpa. Tensile splitting strength ranges from approximately 1.2MPa to 3.0 Mpa.

The tensile strength of masonry in general is relatively small and unreliable (C.o.V. 25% - 40%), especially due to poor unit-mortar bonding. Tensile strength of units, in this case CaSi, is higher than bond strength.

Like for many other brittle materials, tensile strength is relatively small compared to compressive strength. For CaSi the ratio between tensile and compressive strength is between 0.10 and 0.20.

Figure 3 shows the normalized stress strain relationship of calcium silicate based on experiments in earlier projects, e.g. van der Pluijm [9]. The behavior of stony materials under tension and, more specific, the post peak behavior was studied in detail by Van der Pluijm [9] and Hordijk [6]. With the appropriate testing devices fracture energy, i.e. the descending branch in the stress strain diagram, can be established.

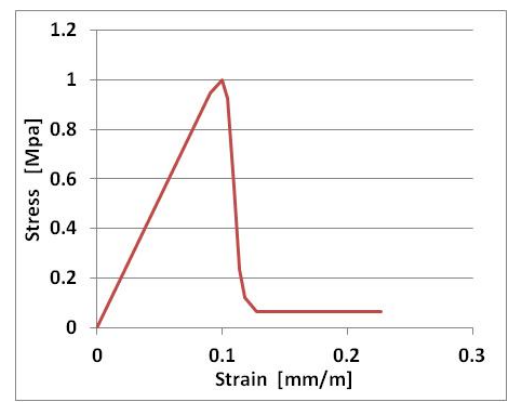

**Figure 4** Example of a stress strain diagram for Calcium Silicate.

# 3 Experiments

## 3.1 Experimental set-up

In this paper the results of experimental research into the effect of the position of the strip is discussed. This parameter covers a relatively wide range of results for the most sensitive but common practical situation. The effect of the out of position placement of the strip on load bearing capacity is probably of concern in design.

Specimens were cut from larger elements as shown schematically in Figure 5. The thickness of the specimen represents a part of the length of the wall. As mentioned earlier, to study the behavior of the joint and the material relatively close to the joint, only a representative part of a wall is needed for experiments. According to the Saint-Venant principle a height of twice width should be sufficient.

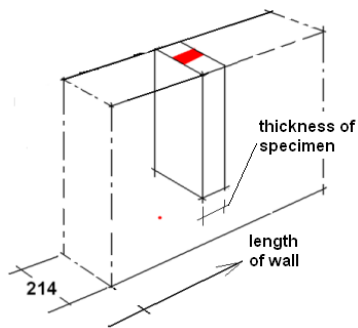

**Figure 5** Cutting scheme of test specimens.

Figure 6 shows the load introduction principle used. Strips, 40 mm wide, were positioned eccentrically from the centre line of the loading machine, indicated in Figure 6 by CL. The bottom of the specimens was a cut surface, the top, with centre strip, was as it would have been in the real building situation.

Rubber strips were cut from larger, 5 mm thick, 80 mm wide rubber strips in the appropriate length and width. The centering strips were put on top of the specimen in the appropriate position. In this way, the building practice situation was simulated.

The block-strip combination was positioned in the machine in such a way that the centre line of the machine coincided with the centre of the strip. This was done to minimize the chance of rotation of the top-load platen.

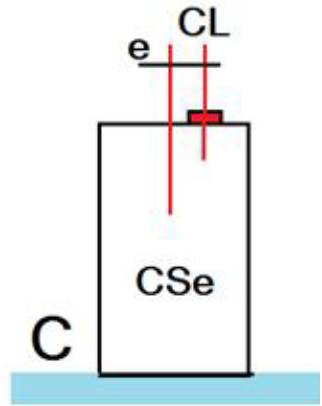

**Figure 6** Load introduction principles. CL = centre line of loading machine.

The specimens were placed on the load platen of the testing machine with their cut surface. The machined surface was on top. A double layer of greased PE foil was positioned between the specimen and the load platen to allow for free lateral movement.

## 3.3 Properties of materials used in the tests

The specimens were made of CS20 elements, a commonly used quality, with a unit compressive strength ($f_b$) of 20MPa and a tensile splitting strength of 2 MPa. Walls made of this quality of units will have a characteristic compressive strength ($f_k$) of 10.2MPa, according to EC6 [13]. The Young's modulus of CS20 masonry is between 8000MPa and 10000MPa, Vermeltfoort [11].

Based on experience, the moisture content of the test-specimens was estimated to be between it 2% en 3% in the given conditions.

Centering strips were made from Styrene butadiene rubber (SBR). Some indicative material properties are: Temperature boundaries range from $-10^o$ to $70^oC$. Tensile strength is 2.5MPa. Strain at fracture equals 150%. Volumetric mass is 1.4g/cm3. Shore hardness is $70^o$+/-5. Maximum compressive strength is 5.0MPa. The manufacturer declared a Young's modulus varying from 10 to 100 MPa; depending on the maximum occurring strain. However, when a piece of rubber is confined between stiffer materials, its behavior will be much stiffer than in a "free" compressive test.

## 3.4 Results

For comparison of results, the specimens overall strength (f'w) and the contact stress (CS) are used. Therefore, the ultimate load is divided by the cross section of the specimen, Equation (3) or the cross section of the strip, Equation (4)

$$f'w = Fult : A \quad (3)$$
$$CS = Fult : As \quad (4)$$

with:
Fult = maximum force observed in the experiment
A = loaded area of the specimen
As = loaded area of the strip, 40x200mm$^2$

Eleven tests were performed, some of which were done twice. Figure 7 shows the load displacement diagrams of five of the tests as an example. It shows the relative brittle failure mode. In Figure 8 the ultimate failure stresses are plotted versus the position of the strip. Figure 9 shows an example of a specimen after testing. A tensile splitting crack is visible.

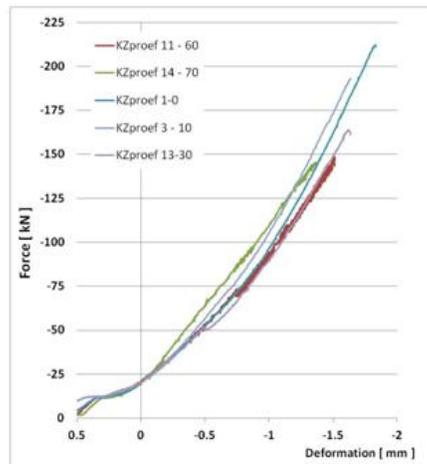

**Figure 7** Load versus displacement of load platen for five example tests.

Main characteristic of the failure of all specimens was the development of a vertical crack, at the edge of the strip (Figure 9A).

A test performed in this way with a centering strip is very similar to the standard splitting test, i.e. a material test. One of the main properties that control failure is the tensile strength. Therefore a relationship between strip-position and tensile splitting strength is expected.

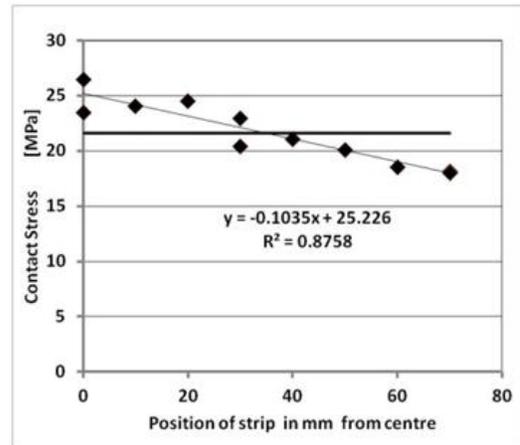

**Figure 8** Contact stress vs strip position.

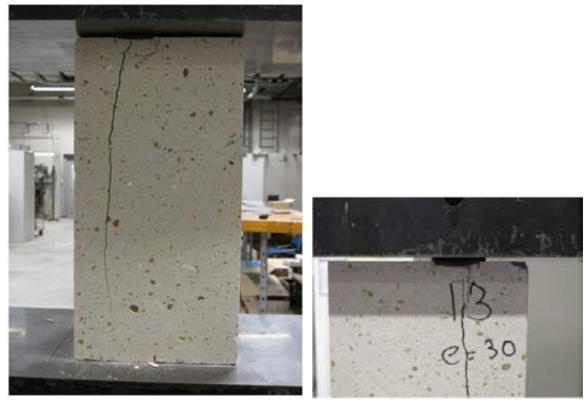

**Figure 9** Splitting of specimen when ultimate load is reached. Right: detail of centering strip

Eccentricity of the load negatively affects the overall wall strength (f'w), as shown in Figure 8, i.e. load bearing capacity of a wall. Buckling effects are usually relatively small for the wall because centering strips are used in cases where the walls are stiff in relation to the floor.

For the eleven performed tests, the following relationship between overall strength (f'w; in MPa) and (c) strip eccentricity (es) with es in mm, was found:

$$f'w = -0.1035\ es + 25.226 \quad (5)$$

The value of $R^2$ was 0.88.
This relationship may depend on dimensions and material used.

## 4 Numerical simulations

The behavior of one of the experiments discussed above was simulated in Comsol. In this paragraph a resume of the results is presented. Two types of simulation were done: one as a sequential linear analysis (SLA) and one with a continuous stress strain model, to study the possibilities to simulate cracking and the effects of variation in material properties.

### 4.1 Stress strain diagram implementation

In the modeling at hand, it was assumed that compressive strength is not critical. Therefore, the behavior under tension was used as failure criterion.

In general, a stress-strain diagram can be described by the initial elasticity modulus ($E_{ini}$), the ultimate stress that was reached (= strength, ft) and the shape of the descending branch.

Especially the descending branch is difficult to establish while it requires dedicated testing control mechanisms (Van Mier [7], Hordijk) [6]. Here it is assumed that the shape of the descending branch is the same in all situations with a relatively smooth behavior around the ultimate load.

Like for all materials, tensile strength and elasticity will randomly vary. For the tension strength, the value ft=2.0MPa was used. This value was obtained with the Brazilian splitting test. A mean value for the modulus of elasticity of 6000MPa was applied. This value was obtained with compressive tests. The two extremes for both strength and E-value results in the four extreme stress strain diagrams, Figure 10.

In this paper, for simplicity, a relatively steep decreasing branch was used in the SLA analyses. In the stress-strain based simulations (next section) a step function for the modulus of elasticity was used.

**Step function**

Using the occurring strain as a measure for failure (i.e. strength) implicates that a relationship between elasticity (E-value) and strain is required. The E-value is the ratio between occurring stress and strain and varies with strain as shown in Figure 10 top. It shows that the E-value is constant in the initial phase of testing, i.e. linear elastic behavior. While fracturing, the E-value decreases fast. Plotting the stress and strain ratio (i.e. E-value) versus occurring strain results in a step-function as shown in Figure 10 bottom.

The step-function for the modulus of elasticity is further normalized as shown in Figure 16. The shape of the step function is assumed to be the same in all simulations.

Values for the initial E ($E_{ini}$) were assigned randomly to areas in the model. These areas measured 20x20 mm on a specimen of 200x400mm$^2$ i.e. 0.1 times the width and 0.05 times the height of the specimen.

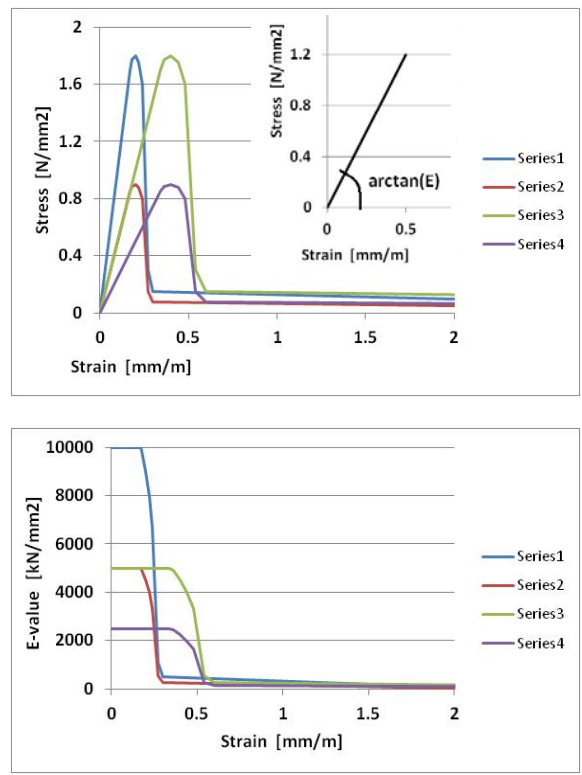

**Figure 10** Four extreme stress strain diagrams and derived strain-modulus of Elasticity relationship. Inset top figure: definition of E; stress=E-value*strain.

### 4.2 Sequential linear analyses

Parameters and values used in the SLA were as follows**.**

The specimen's dimensions were 200mm wide and 300mm high, similar to the dimensions of the specimens in the experiments. The boundary conditions allowed free horizontal movement at the bottom. The rubber centering strip, 70mm wide, positioned 50mm eccentric, was given a vertical displacement.

The material properties of Rubber were as discussed above. A mean E-value of 6000MPa for CaSi was used.

Van de Graaf [10] states: "Numerical analysis of physically nonlinear structural behavior is typically carried out using standard incremental-iterative schemes, but these methods show to have some problems. An alternative way was proposed to analyze this kind of structural behavior and this proposal followed earlier ideas by Van Mier [7] and Beranek and Hobbelman [1] about schematizing masonry compression test specimens using strut and tie models.

The key idea is to start building a model consisting of a number of elements discretized in the usual way using standard finite elements. Then the model is deformed and load distribution in the model is established under assumed linear elastic material behavior. Next, it is determined which element is the most critical; in the model. This means: find the element with the highest tensile stress.

The properties of the element are changed, simply said, the element is removed. For practical reasons, a relatively low E-value is assigned. Then the same procedure is repeated for the reduced model. The process of "finding and removing the critical element" is repeated until the load is below a given threshold or when the deformation becomes too large. Actually, a sequence of models is analyzed under assumed linear elastic behavior, i.e. Sequential Linear Analyses (SLA).

Because removing of an element may give complications and actually changes the appearance of the model, it is in most cases easier to reduce the stiffness considerably.

In terms of applied stress-strain curves this means a fast decrease of stiffness. In the SLE the E-value was reduced to 0.01 times the original value.

Some explorative SLE work was done with constant initial E-values. The main result was finding that the crack pattern developed like in the experiments. Further, the effect of eccentricity of the load on stress and displacements was recognized, Figure 12 and 13 respectively. On: http://www.youtube.com/watch?v=bdz0GM1IzjE the step-wise development of cracks is shown.

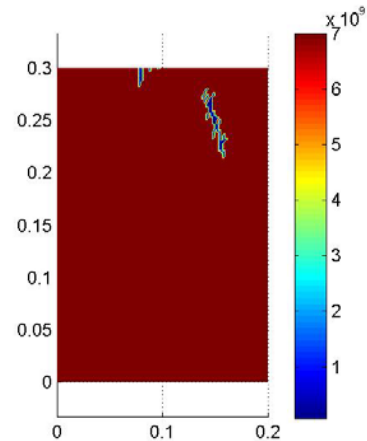

**Figure 11** Places with high and low E-values in the model after hundred steps. Smaller E-values indicate where cracking occurs, comparable with experiments as shown in Figure 9.

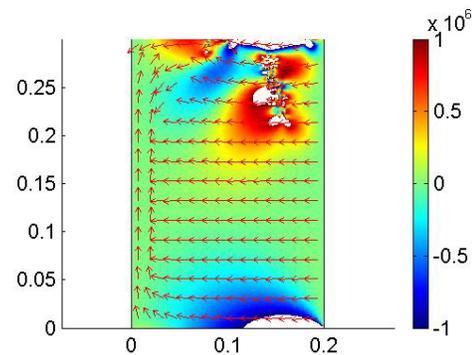

**Figure 12** Stress Distribution over the model after hundred steps. Peak stresses below the corner of the centering strip. Non uniform stress distribution at the bottom due to load eccentricity.

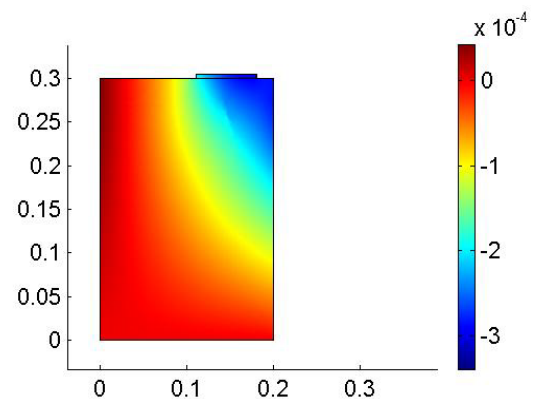

**Figure 13** Displacement contours showing the effect of load eccentricity.

## 4.4 The mechanical Stress-Strain model

In [12] the development of a Heat-Moisture-Stress-Strain (HMSS) model is discussed. In the project at hand, only the stress-strain part was used as described in [4]. The theory is shortly resumed below.

The model was developed as a linear elastic boundary value problem. This type of problem is based on the equilibrium, compatibility and constitutive relationships of a three-dimensional volume element in a continuous body.

The mechanical equilibrium is based on three tensor partial differential equations that can be summarized by the equation of motion such that conditions of static equilibrium are considered:

$$\nabla \sigma + F = 0 \quad (6)$$

Compatibility is expressed by six equations describing the small strain-displacement of a continuous body. The total strain tensor is written as:

$$\varepsilon = \frac{1}{2}[\nabla u + \nabla u^T] \quad (7)$$

The constitutive relationships are based on the linear proportionality between stress and strain tensors. The Voigt notation can be used to express this relationship in matrix form, shown for a two dimensional isotropic problem:

$$\begin{bmatrix} \sigma_{xx} \\ \sigma_{yy} \\ \sigma_{xy} \end{bmatrix} = \begin{bmatrix} C_{11} & C_{12} & C_{13} \\ C_{21} & C_{22} & C_{23} \\ C_{31} & C_{32} & C_{33} \end{bmatrix} \begin{bmatrix} \varepsilon_{xx} \\ \varepsilon_{yy} \\ \varepsilon_{xy} \end{bmatrix} \quad (8)$$

The C-matrix can be further expressed by means of two independent material coefficients, namely Young's modulus, E, and Poisson's ratio, ν. This matrix is shown in Equation (9) for a plane strain problem. In the stress-strain model used, Young's modulus is a step-function of strain and Poisson's ratio is a constant coefficient.

$$C = \frac{E}{(1+\nu)(1-2\nu)} \begin{bmatrix} 1-\nu & \nu & 0 \\ \nu & 1-\nu & 0 \\ 0 & 0 & \frac{(1-2\nu)}{2} \end{bmatrix} \quad (9)$$

This means that stress is related to strain via the E modulus (equation 11). In the thermal analogy:

$$\lambda = f(T) \quad (10)$$

and in applied mechanics

$$E = f(u) \quad \text{or} \quad E = f(\frac{du}{dx}) \quad (11)$$

i.e. $E$ is be a function of $u$ (displacement), or a derivative (= strain, $\varepsilon = du/dx$). In this way, the cumbersome elaboration of "removing" elements like in the SLA is not needed. Further studies are required to investigate whether the iteration process works appropriately.

The theory above implicates that, in the SS model, checks are made for the magnitude of strain and E-values adjusted accordingly.

Actually, the load on the model is applied in steps and while load increases, the E-value for each specific element is smoothly adjusted to the occurring first principal strain in that element, following a step function as shown in Figure 16. The process stops after the requested number of steps is made or when numerical instability occurs.

## 4.5 Simulation Input

In this section some of the main input for the SS model is presented and discussed

Input for size, rubber and boundary conditions was similar to the input in the SLA simulations (section 4.2). Constant values for E, 1000MPa, and Poisson's ratio, 0.45, were assigned to the rubber strip.

Main difference is the way in which cracking is controlled, i.e. strain is governing instead of stress, like in the SLA model. Further, a randomly variable initial E-value and a constant Poisson's ratio of 0.2 were assigned to the CaSi block. Therefore, two interpolation regimes were applied: one for the initial Young's modulus (function name: int1) and one for the step function, similar to the function shown in Figure 10 to simulate fracture (named: PercE). Per area, the initial Young's modulus was a random value between 7700MPa and 6300 MPa, Figure 15.

The (step-)function, shown in Figure 15 is based on the values given in Table 1 and differs from the one shown in Figure 10 which is derived from experiments. On the X-axes, the value of e* is plotted. For practical reasons e* is taken 6000 times the real strain. In the experiments, the strain at ultimate load was 0.333 mm/m. This value may be found by dividing strength by Eini. In figure 15 the E-value starts to decrease by a strain e* = 2 m/m, or 0.333*6000/1000.

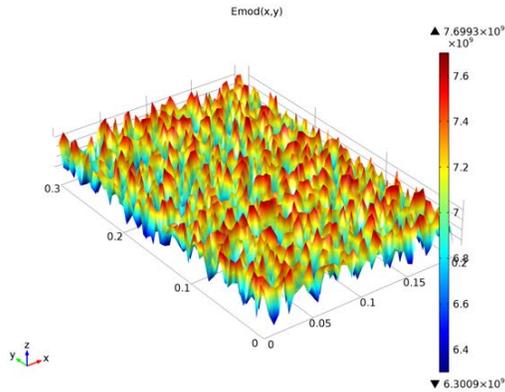

**Figure 14** Initial E-values distribution according to the first interpolation regime (int1).

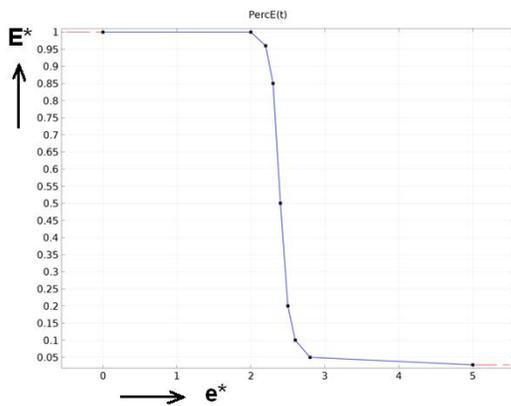

**Figure 15** Step function of E* versus e* (named: PercE) used for the simulation of fracture

Table 1. Strain, E* and stress values used in simulation.

| e* | strain ε | E* | stress MPa |
|---|---|---|---|
|  | mm/m | -- | MPa |
| 0.0 | 0.000 | 1.000 | 0.000 |
| 2.0 | 0.333 | 1.000 | 2.000 |
| 2.2 | 0.367 | 0.960 | 2.112 |
| 2.3 | 0.383 | 0.850 | 1.955 |
| 2.4 | 0.400 | 0.500 | 1.200 |
| 2.5 | 0.417 | 0.200 | 0.500 |
| 2.6 | 0.433 | 0.100 | 0.260 |
| 2.8 | 0.467 | 0.050 | 0.140 |
| 5.0 | 0.833 | 0.028 | 0.140 |
| e*=ε/Eavg; E* = Eini/Eavg; Eavg = 6000MPa | | | |

On the Y-axis the ratio between Eini and the mean value for E (6000MPa) is plotted.

In the simulations, the E-value is a function of the position, x,y and of solid.ep1 as follows:

Exys = 1e7+(PercE(6000*solid.ep1))*Emod(x,y)     (12)

The shape of the descending branch differs slightly from the one shown in Figure 10 and the effects will be studied in subsequent research.

The model load was time dependent; range (0,1,1000). As there were no time dependent parameters, actually 1000 steps were made.

### 4.6 Simulation results

The main results of the SS-simulation are shown in Figures 17, 18 and 19. First principle stresses, Figure 17, show that right hand top corner is most critical. This is confirmed by Figure 18 which shows that in this area the E-values after 778 calculation steps are small compared to the high initial E-values for the rest of the model.

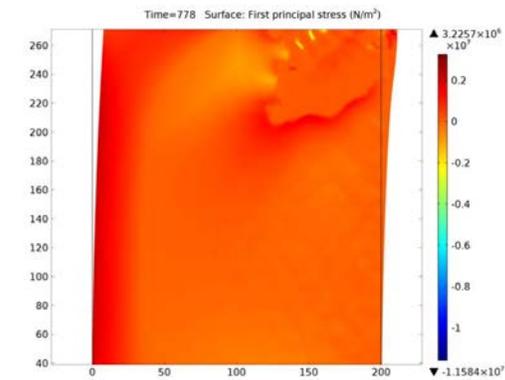

**Figure 16** First principal stresses, compare the right hand top corner with the same area in with Figure 18.

Figure 17 and 18 represent the final phase of simulation. Earlier the development of cracks (i.e. area with low E-values) was visible. A video is given at:
http://www.youtube.com/watch?v=A9gJivBOiNQ

The main simulation result is the load displacement diagram, Figure 19. After a drop in the load, at $1.125*10^7 N/m^2$ the load resumes again. This is different from the load-displacement relationships found in the experiments, Figure 7. Probably, in reality, the fractured specimen differs from the simulation model and consequently not all experimental details are properly taken into account. In reality (the experiment) the test is aborted after the first drop. This first drop is also much larger in the experiments. Further research into this phenomenon is planned.

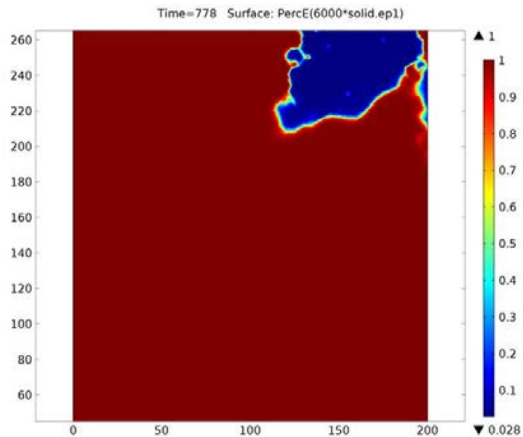

**Figure 17** Distribution of E-values after 778 steps. An area with relatively high initial E-values and an area with low E-values -cracked material- is visble.

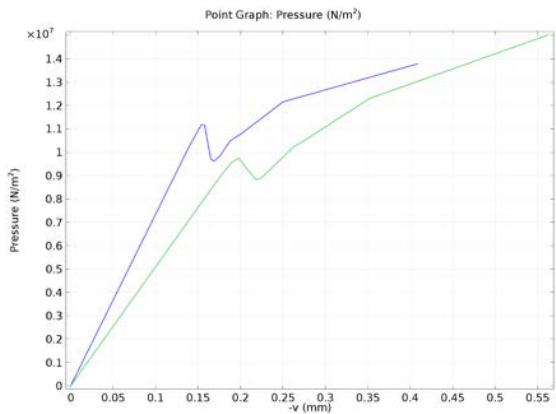

**Figure 18** Load displacement diagram of the two edges of the centering strip, similar to the diagrams from experiments, Figure 7, until the first peak at 1.123 e7 N/m$^2$.

In a next phase of the study, not only E will be randomly assigned to small areas of the model but also strength. That means that e* becomes variable too (see figure 10).

### 4.7 Other aspects.

Some aspects that must be addressed are:
-a the lateral movement of the specimen,
-b the effects of grid refinement
-c the effects of the shape of the step function used, i.e. post peak behavior and fracture energy.
-d whether the first principal strain is the optimal parameter to use.

**ad a**) When a symmetric specimen is loaded symmetrically its deformed shape will be symmetric as well. Bending will not occur. However, when loaded eccentrically the top of the specimen will not only move in vertical but also in horizontal direction, Figure 17. When in a test lateral deformation is partly prevented horizontal reaction forces will develop. For practical reasons no measurements were taken to prevent the development of horizontal forces. It is assumed that these forces would be relatively small due to the deformation properties of the centering strips.

**ad b**) **and c**) Actually, extreme brittle failure is assumed. Which failure criterion will be used also depends on the type of element used and the grid fineness. A simple bar element in a strut and tie model will fail in tension. The load in this type of element is either tension or compression, due to the hinged end conditions. When more complicated elements are used, like beam elements, plate or shell elements, the failure criterion to be used needs some consideration.

**ad d**) Close to the load introduction point confining effects occur. There, the material is in a three dimensional compressive stress state. Consequently, the material seems to be stronger than under the assumed uni-axial stress condition and stress-strain relations are different compared with loading under 3D situations.

## 5 Discussion and Conclusions

This paper shows the possibility to randomly assign strength and E-values to parts of a specimen and to simulate crack patterns.

The continuous stress strain model worked and showed promising results. The direct relation between strain and E-values made it easier to work with than the sequential linear analysis.

In experiments, the main parameter was the position of the strip on the specimen. Therefore, the research will be continued to simulate situations with the strip in the centre and with 20, 40 and 60 mm eccentricity to establish the relation between strip-position and load bearing capacity.

Variation of strength and elasticity over the volume will be addressed in sub sequent work together with some other aspects, like grid refinement and the detailed modeling of post peak behavior.

## 6   References.

1   Beranek, W.J., Hobbelman, G.J., A mechanical model for brittle materials, Proc. of the 9th Int. Brick/Block Masonry Conf., Berlin, Deutsche Gesellschaft fur Mauerwerksbau, Bonn, 694-702, 1991, see also:
http://www.hobbelman.org/gerrie/cv.html
2.  Berkers, W.J.G., Building with calcium silicate elements, Proc. 4th Int. Mas. Conf., London, Vol 1, (7) pp. 176-177, 1995
3   Blauwendraad, J., Plates and FEM, surprises and pittfals, Springer, 2010.
4   Comsol, Structural mechanics module, User's guide, Version 3.3, 2006.
5   HERON, Indirect tensile test on concrete cylinders. Preliminary investigations. Rapport BI-55-19. I.B.C.-T.N.O, 1955. via http://heronjournal.nl/4-3/1.pdf
6   Hordijk, DA, Tensile and tensile fatigue behavior of concrete; experiments modeling and analyses, HERON Vol. 37 no 1, 79 pp. 1992.
7   Mier, J.G.M. van, Fracture processes of concrete, CRC Press, Boca Raton, U.S.A,1997.
8.  Page A.W., Finite Element Model for Masonry Subjected to Concentrated Loads, Journal of Structural Engineering, Vol. 114, No. 8, pp. 1761-1784, 1988.
9   Pluijm Van der, R., Out of plane bending of masonry, PhD Thesis, Eindhoven University of Technology, 1999.
10  Van de Graaf, A, Sequentially linear analysis as an alternative to nonlinear analysis applied to a reinforced glass beam, 7th fib PhD Symp., Stuttgart, 2008
11. Vermeltfoort, A.T. & Ng'Andu, B.M.. Design considerations and the use of CASIELs in medium rise buildings. Proc. of the 3th Int. Conf. on Structural Engineering, Mechanics and Computation (SEMC 2007), Cape Town, South Africa, pp. 1-6 (2007)
12  Williams Portal N.L., Kalagasidis A.S., and Van Schijndel A.W.M., Simulation of heat and moisture induced stress and strain of historic building materials, Proc. of Building Simulation 2011: 12th Conf. of Int. Building Performance Simulation Association, Sydney, 2011. via http://www.ibpsa.org/proceedings/BS2011/P_1130.pdf
13  /…/, EN 1996-1-1, Eurocode 6. Design of masonry structures. General rules for reinforced and unreinforced masonry structures. European com. for standardization, Brussels, 2005.

Video of SS simulation on
http://www.youtube.com/watch?v=A9gJivBOiNQ
Video of SLA simulation on
http://www.youtube.com/watch?v=bdz0GM1IzjE